\documentclass[10pt,letterpaper,twocolumn]{article}

\usepackage{ol2}
\usepackage[draft]{hyperref}
\usepackage{amsmath}
\usepackage{amssymb}

\begin{document}

\twocolumn[ 

\title{Fano resonances in saturable waveguide arrays}

\author{Uta Naether, Daniel E. Rivas, Manuel A. Larenas, Mario I.  Molina, and Rodrigo A. Vicencio}

\address{Departamento de F\'{\i}sica, Facultad de Ciencias, Universidad de Chile, Casilla 653, Santiago, Chile}

\begin{abstract}We study a waveguide array with an embedded nonlinear saturable impurity. We solve the impurity problem in closed form and find the nonlinear localized modes. Next, we consider the scattering of a small-amplitude plane wave by a nonlinear impurity mode, and  discover regions in parameter space where transmission is fully suppressed. We relate these findings with Fano resonances and propose this setup as a mean to control the transport of light across the array.
\end{abstract}


] 


Discrete nonlinear systems have been studied in the context of nonlinear optical waveguide arrays during the last twenty years\cite{rep1,rep2}. One of its main advantages is the possibility to change and control all essential parameters, e.g. geometry, dimensionality, nonlinearity, beam angle, etc. On the other hand, scattering problems have always been of great interest to the physics community due to its capacity to probe the physical nature of many complex systems.  In the context of nonlinear waveguide arrays, the main properties of soliton scattering against impurity potentials have been theoretically\cite{olluis} and experimentally\cite{prlmora} studied, and a rich and complex phenomenology has been observed. During the last years the scattering of plane waves by nonlinear localized modes has opened the possibility to observe Fano resonances\cite{fano} in nonlinear waveguide arrays: Nonlinearity generates several scattering channels which can lead to resonances due to destructive interference and, as a consequence, to total absence of transmission similar to the original Fano problem. As examples we can cite the prediction of Fano resonances in the context of nonlinear quadratic waveguide arrays\cite{x2us} and also, in a very different research area, the recent prediction of similar resonances in Bose-Einstein condensates\cite{prlbec} (see Ref.\cite{miro} as a review of recent findings in this area).  

In the present letter we propose a saturable impurity embedded in a linear waveguide array as a new possible experimental setup to observe Fano resonances. Recently, scattering of plane waves by bright and dark solitons has been experimentally studied in saturable nonlinear media\cite{darkkip} but, to the best of our knowledge, no direct observation of this type of resonances has been implemented yet. We also characterize the main properties of nonlinear impurity modes (NLM) in this type of system and suggest the possibility of a switching-mode scheme, based on a judicious tuning of the system parameters. 

We consider the saturable discrete nonlinear Schr\"odinger (s-DNLS) equation in a dimensionless form with a linear defect and nonlinear interaction only at one site ($n=n_i$) of a 1D waveguide array:
\begin{equation}
-i\frac{\partial u_n}{\partial z} = (u_{n+1}+u_{n-1})+\left(\epsilon- \frac{\beta}{1+|u_n|^2}\right)u_n\delta_{n,n_i}. \label{dnls}
\end{equation}
These coupled-mode equations describe the propagation of light in weakly-coupled waveguides close to the first band of the band-gap structure. Amplitude $u_n$ represents the light amplitude at a guide centered on site $n$, $z$ the propagation coordinate along the waveguides, $\epsilon$ is the strength of the linear defect and $\beta$ the nonlinear coefficient. We include both type of defects, linear and nonlinear, in order to deal with a more general problem. A linear site impurity can be created by altering the geometry of a given waveguide, while at the same time tuning the spacing with its nearest-neighbors in order to keep its coupling with the rest of the array unaltered. On the other hand, the nonlinear response at the impurity site can be boosted by a judicious amount of extra metal doping\cite{stepic}.

\begin{figure}[htb]
\centerline{\includegraphics[width=8.3cm]{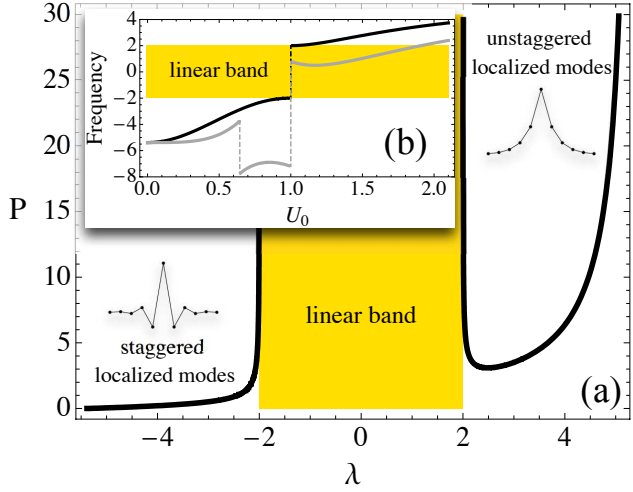}}
\caption{(a) $P$ vs $\lambda$ diagram for NLM. (b) $\lambda$ vs $U_0$ for NLM (black lines) and $\omega_{cc}$ vs $U_0$ for local modes (gray lines).}\label{espe}
\end{figure}

First, we look for stationary localized solutions centered at site $n_i$. We insert into (\ref{dnls}) the ansatz $u_n(z)\equiv U_n(z)=U_0 x^{| n-n_i|} e^{i \lambda z}$ with $U_0, x\in \emph{R}$. $U_0$ is the impurity-mode amplitude and $x$ determines its localization length. By defining $g\equiv \beta/(1+U_{0}^2)$ and imposing $|x| < 1$ we obtain: $x=(g-\epsilon)/2\pm \sqrt{1+((g-\epsilon)/2)^{2}}$ and $\lambda= \pm \sqrt{4+(g-\epsilon)^2}$. The relation between $g$ and $\epsilon$ determines the sign of $\lambda$. For $g > \epsilon$ ($g<\epsilon$) the sign is minus (plus), $x$ and $\lambda<0$ ($x$ and $\lambda>0$), and the solution is staggered (unstaggered). The optical power is defined as $P\equiv\sum_n{|u_n|^2}=U_0^2\sqrt{4+(g-\epsilon)^2}/|g-\epsilon|$. Hereafter, we fix $\beta=10$ and $\epsilon=5$ (Thus, if $U_0=1\rightarrow g=\epsilon$). Figure \ref{espe} shows power vs propagation constant of the impurity mode. From Fig.\ref{espe}(a) we see that there is no power threshold for staggered modes (which are always linearly stable). On the other hand, unstaggered modes posses a threshold that separates two regimes: $\partial P/\partial \lambda<0$ unstable and $\partial P/\partial \lambda>0$ stable solutions. When nonlinear modes approach the linear band ($|g-\epsilon|\rightarrow 0$), the optical power diverges.
%
%
This amplitude-dependent ``transition'' implies that, unlike other impurity type, for a single saturable impurity, one can switch from a staggered to an unstaggered mode just by varying the input power.

Now, we look for Fano resonances, i.e. what would happen, if a small-amplitude plane wave is scattered by a nonlinear impurity mode? In the impurity region, where the interaction takes place, a ``local mode'' can be generated. This mode corresponds to an extra channel, which is opened by the interaction. When this mode is fully excited, a zero transmission of plane waves can occur  due to destructive interference and a Fano resonance appears\cite{miro}.
We assume the plane wave amplitude $\phi_n(z)$ to be much smaller than the impurity-mode amplitude, i.e. $|\phi_0|\ll |U_0|$. We insert $u_n(z) = U_n(z) + \phi_n(z)$ and linearize (\ref{dnls}) with respect to $\phi_n(z)$ obtaining:
\begin{eqnarray}
-i \frac{\partial \phi_n}{\partial z} = \phi_{n+1}+\phi_{n-1}+\left(\epsilon-\frac{g^2}{\beta}\right)\phi_0\delta_{n,n_i}+\nonumber \\
\frac{g^2}{\beta}U_0^2e^{2i\lambda z}\phi_0^*\delta_{n,n_i}.
\label{elm}
\end{eqnarray}
To solve this problem, we use the ansatz
$\phi_n(z)=a_n e^{i\omega z}+b_n^*e^{i(2\lambda-\omega) z}$ in (\ref{elm}), obtaining two coupled discrete equations:
\begin{eqnarray}
\omega a_n=(a_{n+1}+a_{n-1})+\emph{L}a_n\delta_{n,n_i},\label{oc}\\
(2\lambda-\omega )b_n = (b_{n+1}+b_{n-1})+\emph{L}b_n\delta_{n,n_i},\label{cc}
\end{eqnarray}
where $\emph{L}c_n\equiv \left(\epsilon-g^2/\beta\right)c_n+\left(g^2U_0^2/\beta\right) d_n$ with $c_n\neq d_n$ ($c_n,d_n=a_n,b_n$). $a_n$ corresponds to the open channel, i.e. a traveling plane wave with a frequency given by $\omega_k=2\cos k$. $k$ is the plane wave vector which, in a experiment, is related to the input angle. Therefore the linear band covers the region $[-2,2]$. $b_n$ corresponds to the closed channel, whose frequency is determined by the interaction. The resonance occurs when the open channel amplitude at the impurity site is zero and the local mode is fully excited. Therefore, we decouple (\ref{cc}) [taking $a_0=0$] and look for localized solutions of the form $b_n=b_0 y^{|n-n_i|}$ with $|y|<1$. We find that the frequency of this mode is $\omega_{cc}=2\lambda \pm \sqrt{4+\left(\epsilon-g^2/\beta\right)^2}$, where the plus (minus) sign holds for $\epsilon<g^{2}/\beta\ (>g^{2}/\beta)$, and  $\lambda>0\ (<0)$ if $g<\epsilon\ (>\epsilon)$.
We notice that a resonance is only possible when the frequency of the open channel matches the frequency of the closed one. It can be proved that the necessary condition for having $\omega_{cc}$ inside the linear band is $\beta>\epsilon>0$. In any other case $\omega_{cc}$ will be out of the band and no resonances will be observed. This condition also implies that the linear impurity is absolutely necessary for having Fano resonances in the present model. Fig.\ref{espe}(b) shows the existence region of the local nonlinear mode as a function of  $U_0$. The frequency $\omega_{cc}$ lies inside the linear band approximately for $U_0\in [1,1.9]$ and, therefore, only in such a region a plane wave can excite a local mode and be totally reflected.\vspace{-0.5cm}
%
\begin{figure}[h!]
\centerline{\includegraphics[width=7.5cm]{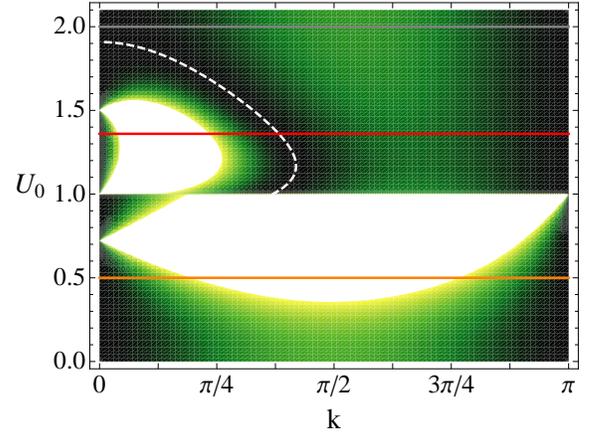}}
\caption{Transmission coefficient $T$ as a function of beam angle $k$ and amplitude $U_{0}$. Bright (dark) regions denote high (low) $T$. Dashed line marks places where $\omega_{k}=\omega_{cc}$ and $T=0$.}\label{theo}
\end{figure}
%

The scattering problem is studied by considering an incoming plane wave and a localized local mode:
\begin{eqnarray*}
a_n=\left\{\begin{array}{c}
B\ e^{ik(n-n_i)}+D\ e^{-ik(n-n_i)}\ ;\ n<n_i \\
F\ e^{ik(n-n_i)} \hspace{2.6cm} ;\ n\geq n_i \\
\end{array}\right.,\label{sca1}
\\ b_n=b_{0}\
y^{|n-n_i|}\ .\hspace{4.78cm}
\label{sca2}
\end{eqnarray*}
By inserting this ansatz in (\ref{oc}) and (\ref{cc}) at sites $n=n_i,n_i\pm1$ we find $F=B+D$
and $\omega_k=2\cos k$. By solving the algebraic problem we get the transmission coefficient $T\equiv \left|F/B\right|^2$ in closed form as $T=4 \sin^2 k/[4 \sin^2 k+\Omega^{2}(k)]$
where,
\begin{equation}
\Omega(k)=\left[\epsilon-\frac{g^2}{\beta}+\frac{g^4 U_0^4/\beta}{\pm2\sqrt{(\lambda-\cos k)^2-1}+\frac{g^2}{\beta}-\epsilon}\right]\ ,\label{T2}
\end{equation}
where the plus (minus) sign is used if $\lambda-\cos k >1\ (<1)$. By fixing $\beta$ and $\epsilon$, the remaining free parameters are the beam angle $k$ and the amplitude $U_0$. Figure  \ref{theo} shows the transmission coefficient in terms of $k$ and $U_{0}$.
First, we clearly see that $T=0$ (Fano resonance) appears only in the range $U_0\sim \{1,1.9\}$ where the frequency of the open channel can coincide with the frequency of the closed channel. Second, resonances are only possible when the NLM is unstaggered ($U_0>1$). This picture suggests a high degree of controllability of the transmission coefficient: By choosing different input powers and/or different input beam angles, we can efficiently control the amount of light going through or reflected back from the impurity region.
\begin{figure}[htb]
\centerline{\includegraphics[width=7.5cm]{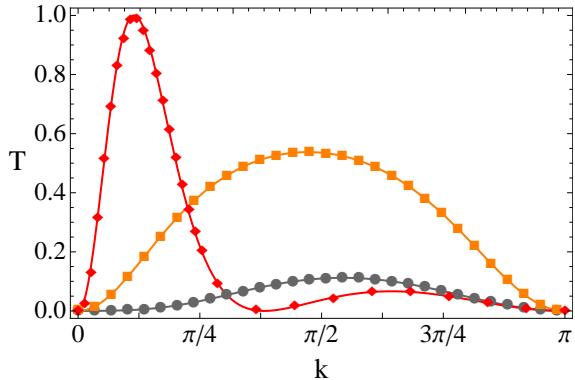}}
\caption{$T$ vs $k$ for $U_0=0.5$ (squares), $1.36$ (diamonds), and $2$ (circles). Lines correspond to theoretical $T$ and symbols to numerical simulations results.}\label{tres}
\end{figure}
%

We perform numerical simulations of (\ref{dnls}) in order to confirm our theoretical findings. As an initial condition, we take a localized nonlinear impurity mode $U_0 x^{| n-n_i|}$ and an incoming plane wave initially centered at $n_0$: $\phi_n=\phi_0 \exp\left[-\alpha(n-n_0)^2\right] \exp\left[ik(n-n_0)\right]$. The initial amplitude $\phi_0=0.01$  was chosen very small compared to $U_0$ to fulfill the analytical criteria. $n_0\ll n_i$ in order to avoid a possible initial overlap. $\alpha=0.001$ provides a wide spacial distribution ($\sim 120$ sites) to correctly simulate the scattering of a ``plane wave'' with a well defined $k$. In Fig.\ref{tres} we show results for three different values of $U_0$.
The agreement between the theoretical $T$ (lines) and direct numerical simulations (symbols) is excellent. This result validates our theory and provides a good support for observing this phenomenon in real experimental setups. Below the critical amplitude $U_0=1$, no resonance exist because no local mode can be excited. The transmission profile is similar to the one found for the scattering of a plane wave against a nonlinear impurity ($U_0=0$) \cite{pla}. For $U_0>1$, the observation of Fano resonances is allowed. However, the stability of the nonlinear mode should also be considered. A small-amplitude plane wave can be viewed as a linear perturbation, therefore the localized solution has to be stable in order to numerically (and experimentally) observe the resonance. For $U_0=1.36$, the red theoretical curve matches perfectly the numerical one (diamonds) where the unstaggered solution is stable. Stability is an extra condition that should be fulfilled (for nonlinear cubic impurities\cite{prlbec}, the nonlinear localized mode is always stable). Finally, for large amplitudes no resonances are possible because $\omega_{cc}$ lies outside the band. The NLM grows in power and becomes an effective wall for the plane wave. Therefore, total absence of transmission is expected for large $U_0$.
%
\begin{figure}[htb]
\centerline{\includegraphics[width=7.5cm]{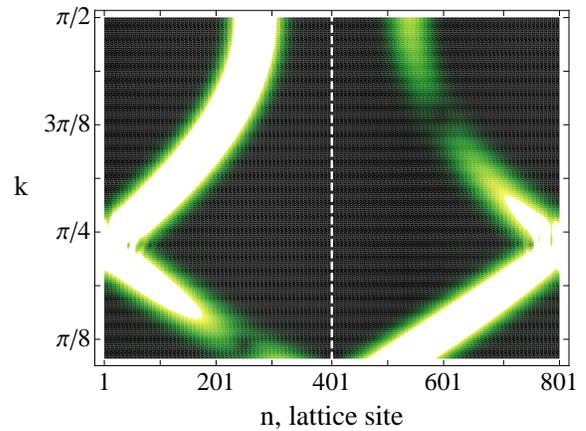}}
\caption{Linear output profile for different $k$. The dashed line marks the position of the impurity site($U_0=1.36$).}\label{out}
\end{figure}
%
Finally, for a fixed $U_0=1.36$ we construct the output profile for different angles [see Fig.\ref{out}]. For small $k$-values, the transmission is very high achieving its maximum ($T=1$) around $\pi/8$. As $k$ is increased, the transmission and reflection coefficients becomes of the same order of magnitude. For $k>\pi/4$, most of the energy is reflected, becoming a maximum exactly at the Fano resonance, where no transmitted light is observed.

In conclusion, we have proposed a new possible setup for observing Fano resonances in optical waveguide arrays. We found a very good agreement between theory and numerical simulations showing that, in principle, this phenomenon could be observed in current experiments. We also showed a switch of nonlinear modes which can be controlled by changing the input power. We believe that our findings constitute a good example of the potential usefulness in using waveguide arrays for all-optical communication systems.

The authors acknowledge financial support from FONDECYT grants 1070897 and 1080374,  and from CONICYT master and doctoral fellowships.


\end{document}